\documentclass[a4paper,11pt]{article}
\usepackage{graphicx,amssymb,
latexsym}
\newcommand{\bea}{\begin{eqnarray}}
\newcommand{\ena}{\end{eqnarray}}

\newcommand{\hs}[1]{\hspace{#1 mm}}

\renewcommand{\c}{\gamma}
\newcommand{\C}{\Gamma}
\renewcommand{\d}{\delta}
\newcommand{\e}{\epsilon}
\newcommand{\s}{\sigma}
\renewcommand{\t}{\theta}
\newcommand{\dsl}{\pa \kern-0.5em /}

\newcommand{\pa}{\partial}

\newcommand{\nn}{\nonumber\\}
\newcommand{\p}[1]{(\ref{#1})}

\begin{document}

\begin{titlepage}

\begin{center}
\hfill UOSTP 5101
\\
\hfill  OU-HET 544
\\
\hfill  {\tt hep-th/0511051}
\vspace{2cm}

{\Large\bf M-theory Supertubes with Three and Four Charges}

\vspace{1.0cm}
{
{\bf Dongsu Bak},$\!^a$ {\bf Kyungyu Kim}$\,^{b}$  and
{\bf Nobuyoshi Ohta}$\,^c$
}

\vspace{0.6cm}

{\it $^a$ Physics Department, University of Seoul,
Seoul 130-743, Korea}

\vskip 0.3cm

{\it $^b$
School of Physics, Seoul National University, Seoul 151-747, Korea
}

\vskip 0.4cm

{\it $^c$ Department of Physics, Osaka University,
Toyonaka, Osaka 560-0043, Japan}

{
({\tt dsbak@mach.uos.ac.kr},
{\tt ohta@phys.sci.osaka-u.ac.jp})
}

\vspace{1.5cm}

{\bf{Abstract}}
\end{center}

Using the covariant M5-brane action, we construct configurations
corresponding to supertubes with three and four
charges. We derive the BPS equations
and study the full structure of the solutions. In particular,
we find new solutions involving arbitrariness in field strengths.

\end{titlepage}
\newpage
\renewcommand{\thefootnote}{\arabic{footnote}}
\setcounter{page}{2}

\section{Introduction}

The counting of microstates associated with the black holes
in superstring theories is a subject of current interest.
The configurations are called BPS solutions which preserve part of
supersymmetry, and this is the key property in the above counting of microstates.
Among others, the BPS configurations called
supertubes~\cite{mateos}-\cite{Shepard:2005wy}
has a dual realization in the brane worldvolume as well as supergravity.
The entropy obtained for the latter is beautifully explained in the microstate
counting in the former realization. This is especially successful in
the supertubes with 2 charges, which is realized as a round D2-brane
configuration of tubular shape involving D0 and F1 on the worldvolume.
(For other related discussions,
see Refs.~\cite{Lunin:2001jy}-\cite{Giusto:2004id}.)

There are also more interesting solutions corresponding to supertubes
with 3 and 4 charges, which have been less
understood~\cite{Bena:2004de}-\cite{Bena:2004td}. The configurations
may have various forms of realization involving three branes in addition
to certain numbers of dipole charges, which may be related by duality
transformations. So it is convenient to consider the configurations
in the context of M-theory.
In this paper, we construct these solutions in the M-theory using the
covariant action of M5-branes~\cite{Covm5} and study their properties.
To discuss this class of solutions, we have to study
the BPS equations whose solutions are the objects of our interest.

The configurations we consider for 3-charge case can be schematically written as
\bea
\begin{array}{|c|ccccccc|}
\hline
\mbox{M2} & 1 & 2 &   &   &   &   &      \\
\mbox{M2} &   &   & 3 & 4 &   &   &      \\
\mbox{M2} &   &   &   &   & 5 & 6 &      \\
\mbox{m5} & 1 & 2 & 3 & 4 &   &   & \t \\
\mbox{m5} & 1 & 2 &   &   & 5 & 6 & \t \\
\mbox{m5} &   &   & 3 & 4 & 5 & 6 & \t \\
\hline
\end{array}
\label{config}
\ena
where M2 stands for M2-branes lying in (12), (34) and (56)
spatial directions within the ``round'' M5-branes in
$(1 2 34 \t)$,
$(1 2 5 6 \t)$,
$(3 4 5 6 \t)$ directions where 1-6 refer to the directions in
the tangent
space. Later we shall find that the shape of $m5$ can be much more
general than the above.
The circular direction of the M5-branes is parametrized by the angle $\t$
in the $(789\natural)$ directions of the target space ($\natural$ stands
for 10).
The total charges of these M5-branes are
zero and for this reason they are denoted as m5.
The M2-branes can be understood as induced by the background fields
on the worldvolume of m5-branes.

If we put one of the M2-brane charges to zero, say the third one,
we could forget about the second and third m5-branes, and get M-theory
realization of supertubes with 2 charges (see for instance \cite{HO}).

Using the covariant action for M5-branes, we make a detailed analysis of
3-charge configurations~\p{config} and briefly discuss the extension to
4-charge case. In section~2, we derive the BPS
equations from the kappa symmetry in the M5-brane action~\cite{Covm5}.
In section~3, we work out the solutions explicitly. Specifically we consider
the case in which the worldvolume of m5 is $T^4$ which is embedded inside
the target space of $T^6$.
This configuration is also discussed in Ref.~\cite{3chrge}, but our
construction is more explicit and contains new  solutions
involving three arbitrary functions of $\t$.
In section~4, we derive the conserved charges of the system and discuss
the bound on the curve imposed by these charges. This should be useful
for the counting of the microstates for fixed conserved charges~\cite{BHO}.
In section~5, we consider the embedding of m5 and M2 system into
a Calabi-Yau (CY) 3-fold instead of $T^6$, and repeat the similar analysis
of the BPS solutions. One of the resulting BPS equations is a nonlinear
instanton equation~\cite{Marino:1999af} whose quadratic fluctuations
for the degeneracy counting and understanding the related fluctuations
is briefly touched upon. In section 6, we briefly discuss
the generalization to the case of four M2-charges in the tangent space of
$T^8$ or CY 4-fold. Section~7 is devoted to conclusion.

\section{BPS equations}

We aim to construct the configuration~\p{config} with the covariant action
of M5-brane proposed in Ref.~\cite{Covm5}, and use the same notation.
In particular, we should note that the flat metric is chosen as
$\eta_{ab}=\mbox{diag}(+1, -1, -1, \ldots)$.

The worldvolume of m5 is parametrized by $(t, \s_i, \t), (i=1,2,3,4)$,
its embedding into the flat target space is given by $(T,Y^a(\s_i)), (a=
1, \ldots, 6)$, and the m5-brane curve in the $(7,8,9,\natural)$ space
is denoted by $X^m(\t), (m=7,8,9,\natural)$. We fix the gauge $t=T$.
The induced vielbein on the worldvolume
\bea
e_i^a = \frac{\pa Y^a}{\pa \s^i},
\label{first}
\ena
relates the gamma matrices as
\bea
\c_i = e_i^a \Gamma_a = \pa_i Y^a \Gamma_a, \quad
\c_0 = \Gamma_0, \quad
\c_5= X^m{}' \Gamma_m,
\ena
and the induced metric
\bea
g_{00} =1, \quad
g_{ij} = \pa_i Y^a \pa_j Y_a, \quad
g_{55}=X^m{}' X_m',
\label{induced}
\ena
where 5 denotes $\t$, the 5-th worldvolume coordinate, and the prime
is its derivative.

Let us now find out the BPS equations. They should be derived from
the kappa symmetry~\cite{Covm5}, which gives
\bea
(\Gamma+1) \e =0,
\label{kappa}
\ena
where
\bea
\Gamma &=& \frac{\sqrt{-g}}{\sqrt{-\mbox{det}(g+ iH)}}
\left[ \frac{\c^{(6)}}{\sqrt{-g}} + \frac{i}{2\sqrt{-(\pa a)^2}} H^{mn} \c_{mn}
\c_p \pa^p a
\right.\nn
&& \hs{20}\left. +\; \frac{\sqrt{-g}}
{8(\pa a)^2} \pa^{m_1} a \e_{m_1 \ldots m_6}
H^{m_2 m_3} H^{m_4 m_5} \c^{m_6}\c_p \pa^p a \right],
\label{kappa1}
\ena
where  $\e^{012345}=-\e_{012345}=1$, and $g$ is the determinant of
the five-dimensional induced metric.

We expect the solution is given by the projection
\bea
\e = P \e_0; \quad
P=P_1 P_2 P_3,
\label{proj}
\ena
where $
P_1 = \frac{1+i\C_{012}}{2},
P_2 = \frac{1+i\C_{034}}{2},
P_3 = \frac{1+i\C_{056}}{2}$
are the projectors.
We choose our gauge for $a$ as
\bea
a=t,
\ena
so that $(\pa a)^2=1$. Our task is now to examine what conditions on the fields
we get from \p{kappa}.

The first term in \p{kappa1} involves
\bea
\c^{(6)} = \c_0 \ldots \c_5
= \C_0 \c_5 \C_{abcd} f^{abcd},
\label{term1}
\ena
where we have defined
\bea
f^{abcd} = \frac{1}{4!} \e^{ijkl} e_i^a e_j^b e_k^c e_l^d .
\label{f4}
\ena
The term~\p{term1} produces two different kinds of terms:
\bea
f^{abcd}\C_{abcd} P = -\frac{1}{2} f^{abcd} \e_{abcdef} \e^{ef}\,P
-3\,
\e^{ab}\left( f^{abce}\e^{ed}-f^{abde}\e^{ec} \right)\C_{cd} P,
\label{term1-0}
\ena
depending on whether $(a,b,c,d)$ belong to two sets out of $(12),(34)$
and $(56)$, or $c$ and $d$ belong to different sets other than $(a,b)$.
Here we have used the projection condition~\p{proj} and
defined $\e^{ab}$ as $6\times 6$ matrix:
\bea
(\e)^{ab} \equiv \left(
\begin{array}{ccc}
I & 0 & 0 \\
0 & I & 0 \\
0 & 0 & I
\end{array} \right);
I \equiv \left(
\begin{array}{cc}
0 & 1 \\
-1 & 0
\end{array} \right).
\ena

The second term in~\p{kappa1} is decomposed as
\bea
\frac{1}{2} \left[ 2 f^{(05)}\c_0 \c_5 + 2 f^{(0)a}\c_0 \C_a
+ 2 f^{(5)a}\c_5 \C_a + f^{ab} \C_{ab} \right] \c_0,
\label{second}
\ena
where we have defined
\bea
f^{ab} = H^{ij} e_i^a e_j^b, \quad
f^{(0)a} = H^{0i} e_i^a, \quad
f^{(5)a} = H^{5i} e_i^a, \quad
f^{(05)} = H^{05}.
\ena
The first three terms in eq.~\p{second} are rewritten as
\bea
-\c_5 f^{(05)}
-\C_a f^{(0)a}
+\C_0 \c_5 \C_a f^{(5)a},
\label{term2-1}
\ena
and the last term yields
\bea
-\frac{i}{2} \e_{ab} f^{ab}P +\frac{1}{4} \C_0 \C_{ab} (f^{ac}\e^{cb}-
f^{bc}\e^{ca} )P\,.
\label{term2-2}
\ena

Finally we come to the third term in~\p{kappa1}. We find that they give
\bea
&& -\frac{\sqrt{-g}}{8}\left[ \e_{ijkl} H^{ij} H^{kl} g^{55} \c_5
+ 4 \e_{ijkl} H^{jk} H^{l5} \c^i \right] \C^0 \nn
&=& -\frac{\sqrt{-g}}{8}\left[ |g^{55}| \e_{ijkl} H^{ij} H^{kl} \C_0 \c_5
+ 4 \e_{ijkl} H^{jk} H^{l5} g^{im} e_m^a
\C_a\C_0 \right] \nn
&=& -\frac{\sqrt{-g}}{8}\left[ |g^{55}| \e_{ijkl} H^{ij} H^{kl} \C_0 \c_5
-i 4 \e_{ijkl} H^{jk} H^{l5} g^{im} e_m^a
\C_b\e_{ab} \right].
\label{term3}
\ena

We have the $\c_5$ term only in the first term in \p{term2-1} to obtain
\bea
f^{(05)}=0,
\label{bps0}
\ena
and $\C_0 \c_5$ terms only from the first term in \p{term1-0} and \p{term3}, so
\bea
\frac{1}{2} \e_{abcdef} f^{abcd}\e^{ef}
+ \frac{|g|}{8} |g^{55}| \e_{ijkl} H^{ij} H^{kl} =0.
\label{bps1}
\ena
The second term in \p{term1-0} is the only $\c_5 \C_{cd}$ term, so we must have
\bea
\e^{ab}\left(
f^{abce}\e^{ed}-f^{abde}\e^{ec}
\right) = 0.
\label{bps2}
\ena
The second term in \p{term2-1} and the last term in \p{term3} involving $\C_a$ give
\bea
f^{(0)a}-\frac{i}{2}\sqrt{-g} \e_{ijkl} H^{jk} H^{l5} g^{im}
 e_m^b 
\e_b{}^a=0.
\label{bps3}
\ena
The third term in \p{term2-1} gives
\bea
f^{(5)a}=0\,,
\label{bps4}
\ena
and so $H^{l5}=0$. Combined with \p{bps3}, this gives
\bea
f^{(0)a}=0\,.
\label{bpskk}
\ena

The last term in \p{term2-2} gives
\bea
f^{ac}\e^{cb}-
f^{bc}\e^{ca}=0,
\label{bps5}
\ena
Finally the first term in \p{term2-2} should cancel 1:
\bea
-\frac{i}{2} \e_{ab} f^{ab} + \frac{\sqrt{-\mbox{det}(g+iH)}}{\sqrt{-g}}=0.
\label{bps6}
\ena

We have thus derived all the BPS equations resulting from the kappa symmetry.
They are eqs.~\p{bps0} -- \p{bps6}.
We are now going to find solutions to these equations.

\section{BPS solutions}

To solve our BPS equations, we first make further gauge choice
\bea
(Y^1, Y^2, Y^3, Y^4) = ( \s^1, \s^2, \s^3, \s^4).
\label{gauge1}
\ena

Let us first examine \p{bps2}. Defining the complex coordinates
\bea
Z=Y^5+iY^6; \quad
w_1= \s^1 +i \s^2, \quad
w_2=\s^3 +i \s^4,
\label{compc}
\ena
and writing explicitly the indices, we find
\bea
f^{1,2,3+i4,5+i6} = f^{3,4,1+i2,5+i6} = f^{5,6,1+i2,3+i4}=0.
\label{bps21}
\ena
Using the definition~\p{f4}, we find that \p{bps21} is equivalent to
\bea
\pa_{\bar w_1} Z = \pa_{\bar w_2} Z = 0.
\label{holo}
\ena
(The last equality in \p{bps21} follows from these relations.)

Next \p{bps5} can be rewritten as
$
H^{w_1 w_2}=
H^{w_1 j} \pa_j Z=
H^{w_2 j}\pa_j Z=0.
$
The latter two conditions follow from (\ref{holo}) and
$H^{w_1 w_2}=0$, so the first condition
\bea
H^{w_1 w_2}=0,
\label{bps51}
\ena
is the only relevant relation. We note that this can be also written as
\bea
H^{13}=H^{24}, \quad
H^{14}=H^{32}.
\label{bps52}
\ena

On the other hand, \p{bps0}, \p{bps4} and \p{bpskk} mean that
\bea
H^{50}=H^{5i}=H^{0i}=0.
\label{bps31}
\ena

\subsection{General analysis of BPS equations}

To examine the remaining conditions~\p{bps1} and \p{bps6}, we need
the induced metric~\p{induced}:
\bea
g_{ij}=-\delta_{ij} - \frac{1}{2}(\pa_i Z \pa_j \bar Z + \pa_j Z \pa_i \bar Z).
\label{metexp}
\ena
With the help of (\ref{holo}), one may show that
$g_{12}=0$, $g_{34}=0$ and
\bea
g_{11}= g_{22}\,, \ \  g_{33}= g_{44} \,, \ \ g_{13}= g_{24}\,, \ \
g_{14}= -g_{23}\,.
\ena
Also by a direct computation, it is straightforward to show that
\bea
g_{(4)}={\rm det} g_{ij}= (g_{11} g_{33}-g^2_{13}-g^2_{23})^2
=(1+\nabla Z \cdot \nabla \bar Z /2)^2\,.
\label{metdet}
\ena
Then \p{bps1} can be  cast into
\bea
\frac{1}{8}\sqrt{g_{(4)}}\,\, \e_{ijkl}H^{ij} H^{kl}
= -1.
\label{bps11}
\ena

One also has
\bea
-\mbox{det}(g+iH)= |g_{55}|g_{(4)}\Big[ 1-\frac{1}{2} H_{ij} H^{ij}
+ \frac{1}{8^2} (\sqrt{g_{(4)}}\epsilon_{ijkl} H_{ij} H_{kl} )^2 \Big]\,,
\ena
which may be rearranged to
\bea
-\frac{\mbox{det}(g+iH)}{|g|}= \left(1+\frac{\sqrt{g_{(4)}}}{8}\,\,\e_{ijkl}
H^{ij} H^{kl}\right)^2
- \left(
\frac{1}{2} H_{ij} H^{ij} +\frac{\sqrt{g_{(4)}}}{4}\,\e_{ijkl}H^{ij} H^{kl}\right).
\label{38}
\ena
Using the properties of $g_{ij}$,
the term in the second parenthesis may be written as
\bea
&& \{H^{12} g_{11}\!+\!H^{34} g_{33} -(H^{13}\!+\!H^{24})g_{14}
\!+\!(H^{14}\!+\!H^{32})g_{13} \}^2\nonumber\\
&& +\; \{(H^{13}+H^{42})^2+ (H^{14}+H^{23})^2 \}
(g_{11} g_{33}-g_{13}^2-g^2_{14}),
\ena
whose second term vanishes due to \p{bps52}. Since the first term in \p{38}
is zero due to \p{bps11}, the BPS equation~\p{bps6} reduces to
\bea
\frac{\sqrt{-\mbox{det}(g+iH)}}{\sqrt{-g}}= 
-i (H^{12}g_{11}
+ H^{34}g_{33}
-2H^{13}g_{14}
+2H^{14}g_{13}
)=  \frac{i}{2} \e^{ab} f^{ab}
\,.
\label{bps7}
\ena
Using \p{holo} and \p{metexp}, it is easy to check that this is automatically
satisfied.

To summarize, our BPS equations are eqs.~\p{holo}, \p{bps52}, \p{bps31}
and \p{bps11} which can be written as
\bea
\left(1+{|\nabla Z|^2\over 2}\right)(H^{12}H^{34} -(H^{13})^2-(H^{23})^2)
= -1\,.
\label{bpsf3}
\ena

There is still an equation one has to satisfy~\cite{Covm52,Sorokin:1997ps}:
\bea
H_{mnp}\partial^p a = V_{mn},
\label{gauss}
\ena
where $V_{mn}$ is defined by
\bea
V_{mn}=-2 {\sqrt{-(\partial a)^2}\over \sqrt{-g}}{\partial{\sqrt{-(g+iH)}}\over
\partial H^{mn}}.
\ena
Also $H^{mn}$ is defined by
\bea
H^{mn}={1\over 3! \sqrt{-g}\sqrt{-(\pa a)^2}}\e^{mnlpqr}H_{pqr}\partial_l a,
\label{hmn}
\ena
from which the equation, $H^{i0}=H^{50}=0$, follows with the gauge
choice $a=t$.

In the present case, one has
\bea
H^{ij}= -{i\over 2 \sqrt{-g}} \e^{ijkl} H_{kl5},
\label{hij}
\ena
and $V^{ij}$ is given by
\bea
V_{ij}=  {i\sqrt{-g}\over \sqrt{-(g+iH)}}\left(
H_{ij}+{1 \over 2} \sqrt{g_{(4)}}\, \e_{ijkl} H^{kl}\right).
\ena
Using \p{bps7}, one finds
\bea
F_{ij}\equiv H_{ij0} = {\sqrt{g_{(4)}}(h_{ij} + {1 \over 2} \sqrt{g_{(4)}}
\, \e_{ijkl} h^{kl}) \over \left( g_{11} h_{34} + g_{33} h_{12}
+ 2 g_{14} h_{13}+ 2 g_{13} h_{41} \right) },
\label{fij}
\ena
where we have defined
\bea
h_{ij}= H_{ij5}/\sqrt{|g_{55}|}.
\ena
{}From this and \p{bps52}, one may show that
\bea
&& F_{12}= g_{11}\,, \ \ F_{34}=g_{33}, \nonumber\\
&& F_{13}=F_{24}= - g_{14}
\,, \ \ F_{14}=F_{32}= g_{13}.
\label{Fij}
\ena
This may be succinctly written as
\bea
F_{ij}={1\over 2} (\e_{ik} g_{kj}-\e_{jk} g_{ki} ),
\label{gausssol}
\ena
which is proportional to the K\"ahler two form.
Introducing the K\"ahler form $k= g_{p\bar{q}} dw^p \wedge d\bar{w}^{\bar{q}}$,
we get $F= F_{p\bar{q}} dw^p \wedge d\bar{w}^{\bar{q}}=i k$ where
$p,q,r=1,2$ and $\bar{p},\bar{q},\bar{r}=1,2$ denoting respectively
the holomorphic and antiholomorphic indices of the m5 worldvolume.
The Bianchi identity
is satisfied since the K\"ahler form is closed, i.e.  $dk=0$.

In addition, using \p{bps52}, one may show that
\bea
h_{13}=h_{24}\,, \ \  h_{14}=h_{32}\,.
\label{self}
\ena

Finally, one may consider  more general ansatz where $Y^a$'s are also
dependent on $\theta$, i.e.  $Y^a(\sigma_i,\theta)$ and $X^m(\theta)$.
The induced metric in the 5 space including the $\theta$ direction
may be written as
\bea
ds^2_{(5)}= g_{ij} d\sigma^i d\sigma^j + 2 g_{i5}d\sigma^i d\theta
+g_{55} d\theta^2
\ena
where
\bea
g_{ij}= \partial_i Y^a \partial_j Y_a\,,  \ \ \
g_{i5}= \partial_i Y^a  Y'_a\,, \ \ \ g_{55}= {X^m}'X'_m +{Y^a}'  Y'_a\,.
\ena
The similar analysis of  BPS equations goes through in this case too.
With the gauge choice of (\ref{gauge1}), $Z$ should be holomorphic function of
$w^p$ and  $H^{w^1 w^2}=0$ with  $H^{0i}=H^{05}=H^{i5}=0$.
Considering the three form,  $H_{ijk}=0$ follows from the relation $H^{i5}=0$.
We introduce a two form
in the four space by $\tilde{h}_{ij}= H_{ij5}$. Then $d\tilde{h}=0$
due to the Bianchi identity of the three form field.

Eq.~(\ref{gauss}) implies
$F_{ij0}\,d\sigma^i\wedge d\sigma^j =i g_{p\bar{q}}\, dw^p\wedge
d\bar{w}^{\bar{q}}$, $F_{p50}= i g_{p5}$ and $F_{\bar{p}50}=-ig_{\bar{p}5}$.
With these expressions,
the Bianchi identity for the three form field can be checked.
 The final BPS equation is given by
\bea
{1\over 8}\, \epsilon^{ijkl}\, \tilde{h}_{ij}
\tilde{h}_{kl}= G_{55}\sqrt{g_{(4)}}\,,
\ena
where $G_{55}=|g_{55}- g^{ij} g_{i5} g_{j5}|$.  Though interesting,
we will not analyze this generalization further in this note.

\subsection{A simple case}

Let us pause to discuss the simple case of $Z=0$. We have $g_{ij}=-\delta_{ij}$.
{}From the BPS equations, there are following set of simple solutions.
By eq.~\p{self}, we set
\bea
h_{13}=h_{24}=a(\theta)\,, \ \ \
h_{14}= h_{32}= b(\theta)\,, \ \ \
h_{12}=c(\theta),
\label{exam}
\ena
where $a$, $b$ and $c$ are arbitrary function of $\theta$ only.
Noting that $H^{12}=-ih_{34}, H^{34}=-ih_{12}, H^{13}=ih_{24}$ and
$H^{23}=-ih_{14}$, we find from \p{bpsf3}
\bea
h_{34}= (1+a^2+b^2)/c\,.
\label{exam1}
\ena
Further, one finds
$F_{12}=F_{34}=-1$ and $F_{13}=F_{24}=F_{14}=F_{32}=0.$
This is a good solution for the case where 123456 directions are
compactified on $T^6$.

\subsection{New solutions}

Using \p{metdet} in \p{hij}, we have
\bea
H^{12} = \frac{-i}{1+\frac{1}{2}|\nabla Z|^2} h_{34}, &&
H^{34} = \frac{-i}{1+\frac{1}{2}|\nabla Z|^2} h_{12}, \nn
H^{13} = H^{24} = \frac{i}{1+\frac{1}{2}|\nabla Z|^2} h_{24}, &&
H^{23} = -H^{14} = \frac{-i}{1+\frac{1}{2}|\nabla Z|^2} h_{14}.
\ena
Combined with \p{bpsf3}, this yields
\bea
h_{34} h_{12} - h_{24}^2 - h_{14}^2 = 1+\frac{1}{2} |\nabla Z|^2.
\label{final}
\ena

One has to impose various condition on the field strengths following from
the  Bianchi identity. It is
\bea
\e^{ijkl}\partial_j  h_{kl}=0\,. 
\ena
Introducing the potential $b_i$ by
$h_{ij}=\partial_i b_j - \partial_j b_i $, eq.~(\ref{self})
may be rewritten as
\bea
h_{13}- h_{24}+i(h_{14}+h_{23})=
(\partial_1+i\partial_2) (b_3+i b_4)-
(\partial_3+i\partial_4) (b_1+i b_2)=0.
\ena
The general solutions are given by
\bea
b_1+i b_2= (\partial_1+i\partial_2)(G_R+iG_I)\,,\ \
b_3+i b_4= (\partial_3+i\partial_4)(G_R+iG_I) \,.
\ena
Choosing the gauge $G_R=0$ and introducing $G=-G_I$, this is written as
\bea
b_i=\e^{ij}\partial_j G\,.
\label{pot}
\ena
One then finds
\bea
&& h_{12}=-(\partial_1^2+\partial_2^2)G\,, \ \
h_{34}=-(\partial_3^2+\partial_4^2)G\,, \nn
&& h_{13}=(\partial_1\partial_4-\partial_2\partial_3)G\,, \ \
h_{14}=-(\partial_1\partial_3+\partial_2\partial_4)
G\,.
\ena

Similarly one has
\bea
a_i= \e^{ij}\partial_j K 
\
\ena
for $F_{ij}=\partial_i a_j - \partial_j a_i $.
It is easy to show that $K$ giving \p{Fij} is
\bea
K
= {1\over 4} (|w_1|^2+ |w_2|^2 + |Z|^2)
\ena
which is proportional to the K\"ahler potential.

Since (\ref{final}) may be written as
\bea
h_{34} h_{12} - h_{24}^2 - h_{14}^2 =
F_{34} F_{12} - F_{13}^2 - F_{14}^2,
\label{final1}
\ena
there is a trivial solution
$G=\pm K$. 
This is the solution presented in Ref.~\cite{3chrge},
which does not include any arbitrariness in the field strengths.
We would like to examine whether there exist other independent solutions.

The solutions~\p{exam} and \p{exam1}
of the $Z=0$ case involve three arbitrary functions of $\theta$.
We expect that the similar type of solutions exist
for the nonvanishing $Z$.
We shall find new solutions involving the
arbitrariness in the field strengths.

Let us now present our new solutions that involve arbitrary functions.
Consider the case where
$Z= f_1(w_1)+ f_2(w_2)$ where $f_a$ are arbitrary holomorphic
functions that depend only on $w_a$. For this, the metric
becomes
\bea
&& g_{11}= -1-f'_1\bar f'_1\,,\ \  g_{33}= -1-f'_2\bar f'_2, \nn
&& g_{13}=-{1\over 2}(f'_1\bar f'_2+f'_2\bar f'_1)\,, \ \
g_{14}={i\over 2}(f'_1\bar f'_2-f'_2\bar f'_1).
\ena
The solution involving arbitrary functions of $\theta$ then becomes
\bea
&& h_{12}=-c(\theta)(1+f'_1\bar f'_1)\ \,,\ \
h_{34}= -{1+f'_2\bar f'_2\over c(\theta)} , \nn
&& h_{14} = -\frac{e^{ia(\t)}}{2} f'_1\bar f'_2
-\frac{e^{-ia(\t)}}{2} f'_2\bar f'_1 \,,\nonumber\\
&& h_{13} = -i \frac{e^{ia(\t)}}{2} f'_1\bar f'_2
+i \frac{e^{-ia(\t)}}{2} f'_2\bar f'_1 \ ,
\ena
One can check that the above satisfy (\ref{final1}) and the Bianchi identity.
In fact, we find that the gauge potential~\p{pot} giving these field strengths is
\bea
G = \frac{c(\t)}{4}(|w_1|^2 +|f_1|^2) + \frac{|w_2|^2 +|f_2|^2}{4 c(\t)}
+ \frac{e^{ia(\t)}}{4} f_1 \bar f_2
+ \frac{e^{-ia(\t)}}{4} \bar f_1 f_2.
\ena

To complete the analysis, one may consider a small fluctuation around
the solution $h_{ij}={1\over 2} (\e_{ik} g_{kj}-\e_{jk} g_{ki} )$.
The small perturbation equation becomes
\bea
&& (1+ f'_1\bar f'_1) \partial_{w_2} \partial_{\bar{w}_2}\delta G
+ (1+ f'_2\bar f'_2) \partial_{w_1} \partial_{\bar{w}_1}\delta G \nonumber\\
&& - f'_1\bar f'_2 \; \partial_{w_2} \partial_{\bar{w}_1}\delta G
- f'_2\bar f'_1 \; \partial_{w_1} \pa_{\bar{w}_2}\delta G =0.
\label{zero}
\ena
One can check the above deformation by $a$ and $c$ gives two independent
solutions to these zero-mode equations. To check if this exhaust all
possibility, let us find out the number of zero modes.

\subsection{Zero-mode equations}

We first consider the simplest case where $f_1=f_2=0\; (Z=0)$. The zero-mode
equation~\p{zero} becomes the Laplace equation in the flat four dimensions:
\bea
(\partial_1^2+ \partial_2^2+ \partial_3^2+ \partial_4^2)\delta G=0\,,
\ena
which leads to the equation
\bea
(\partial_1^2+ \partial_2^2+ \partial_3^2+ \partial_4^2)\delta h_{ij}=0\,,
\ena
together with the Bianchi identity.

For the case of $T^4$, $\d h_{ij}$ defined on $T^4$ should have the
absolute minimum, which contradicts the properties of Laplace equation
unless $\delta h_{ij}$'s are constant functions.

For the case of flat four plane, again $\delta h_{ij}$ should be
constant if one does not allow singularity or divergences at infinity.
Then one may show that there are only three independent
regular solutions,
\bea
&& \delta h_{12}=-\delta h_{34}=\delta c(\theta), \nonumber\\
&& \delta h_{13}= \delta h_{24}= \delta a(\theta), \nonumber\\
&& \delta h_{14}= \delta h_{32}= \delta b(\theta),
\label{ex1}
\ena
where $\delta a$,   $\delta b$, and  $\delta c$ are functions of
$\t$ only. This is consistent with the full solution in (\ref{exam}).

Let us consider the general compact four K\"ahler
manifold with a K\"ahler metric $g_{p\bar{q}}$.
The equation we have to solve is the Hermitian Yang-Mills equation
\bea
g^{p\bar{q}} \delta h_{p\bar{q}}=0,
\label{her}
\ena
with $\delta h_{pq}=\delta h_{\bar{p}\bar{q}}=0$.
Of course  the Bianchi identity
\bea
d \, \delta h = 0,
\ena
has to be satisfied.
Note that the Hermitian Yang-Mills equation combined with the
Bianchi identity implies that
\bea
d^\dagger \delta h = 0\,,
\ena
which follows from $g^{p\bar{q}} \nabla_p \delta h_{\bar{q} r} =
\nabla_r(g^{p\bar{q}} \delta h_{p\bar{q}})$.
Therefore $\delta h$ is a harmonic (1,1) form which satisfies
an additional constraint of the Hermitian Yang-Mills equation
(\ref{her}). From the above computation, the harmonic (1,1) form
satisfies in general
\bea
g^{p\bar{q}} \delta h_{p\bar{q}}= C,
\ena
where $C$ is constant. Therefore the number of zero mode, $I$,
in the compact K\"ahler space is given by
\bea
I= h_{1,1}-1,
\ena
where $h_{p,q}$ denotes the Hodge number of the harmonic
$(p,q)$ form. For $Z=0$, the four manifold describes $T^4$, whose
Hodge number $h_{1,1}$ is four. Thus for $Z=0$, $I=3$, which is
consistent with the explicit construction (\ref{ex1}).

For $S^2 \times S^2$ which is K\"ahler, $h_{1,1}=2$ and, hence,
$I=1$. More explicitly, let us use the complex coordinate $w^1$ and $w^2$
for the spheres. Then the harmonics forms are given by
the volume forms $\sqrt{g_{(1)}} dw^1 \wedge d\bar{w}^1$ and
$\sqrt{g_{(2)}} dw^2 \wedge d\bar{w}^2$. The hermitian Yang-Mills equations
are satisfied only by
\bea
\delta h =
\sqrt{g_{(1)}} dw^1 \wedge d\bar{w}^1 - \sqrt{g_{(2)}} dw^2 \wedge
d\bar{w}^2\,.
\ena
Therefore we find that $I=1$ by the explicit construction.

\section{Charges and Conserved Quantities}

In order to understand what microstates are possible, we should look at
the configurations with various conserved quantities fixed~\cite{BHKO}.
In this section, we derive explicit expressions of these quantities and
examine bounds on the shapes of the BPS configurations.

{}From the Chern-Simons term,
\bea
\frac{1}{2}\int H \wedge  C_{(3)},
\ena
the M2-brane charge density extended in the $(ij)$
direction can be identified as\footnote{The actual charge here is
 twice of the above   Chern-Simons contribution\cite{Sorokin:1997ps}.}
\bea
Q_{ij}={1\over 4\pi} \e_{ijkl} \int^{2\pi}_0 d\theta
H_{kl5}= {1\over 4\pi} \e_{ijkl} \int^{2\pi}_0 d\theta h_{kl}\sqrt{|g_{55}|}
\ena
where the density is the charges per unit coordinate area in the
$(kl)$ directions\footnote{Here we consider the holomorphic four submanifold
independent of $\theta$ for simplicity.}.
To compute the angular momentum, let us first compute the linear
momentum density in the $\theta$ direction. This is given by
\bea
{\cal P}_5 = \Pi^{0ij} H_{5ij},
\ena
where the displacement $\Pi^{0ij}$ is defined by
\bea
\Pi^{0ij}={\partial {\cal L}\over \partial \dot{A_{ij}}}=
{\sqrt{-g}\over 2} H^{*0ij}={1\over 4} \e^{ijkl} H_{kl 5}=
{i\sqrt{-g}\over 2}H^{ij}
\,.
\ena
For the momentum density,
we ignore the contributions from other fields or time dependence,
which vanishes due to the BPS conditions. Using the expression
for the displacement and \p{bps11}, the momentum density is evaluated as
\bea
{\cal P}_5=-\sqrt{g_{(4)}}|g_{55}|\, {\sqrt{g_{(4)}}\over 8}
\e_{ijkl} H^{ij}H^{kl}
= \sqrt{g_4}X'_m X'_m .
\ena
In the target space viewpoint, the momentum density along the circle direction
of tube, is then given by
\bea
P_m=
\sqrt{g_{(4)}}\,
X'_m 
\,.
\ena
The angular-momentum density
in the transverse space is given by
\bea
J_{mn}= {\sqrt{g_{(4)}}\over \pi} \int_R dX^m\wedge d X^n\,.
\ena
where $R$ denotes the region enclosed by the arbitrary
curve ${\cal C}$ of $X^m(\theta)$.

Since the angular momentum is proportional to the area, the length $L$
of the cross section of the supertube
\bea
L={1\over 2\pi} \int^{2\pi}_0 d\theta \sqrt{X'_m X'_m}\,,
\ena
is bounded from below by the square root of the area~\cite{BHO}.
When the curve is confined in the (12) space, one has the bound
\bea
|J_{12}| \le \sqrt{g_{(4)}}\, L^2\,.
\ena
Considering more general curve, one may get
\bea
|J'_{12}|+|J'_{34}| \le \sqrt{g_{(4)}}\,  L^2\,,
\ena
where
$J'_{12}$ and $J'_{34}$ represents the value of Cartans
when only Cartans remain nonvanishing by the $SO(4)$
rotations.

The curve is also constrained by the combination of charges. The
inequality is given by
\bea
\sqrt{g_{(4)}}\, L^2 \le W(Q),
\ena
where
\bea
W(Q)=
{1\over 8}|\e^{ijkl} Q_{ij}Q_{kl}|
\,.
\ena
The proof is as follows.
Let us work in the local Lorentz frame where $Q_{ij}$
is block diagonalized with only nonvanishing components $Q_{12}$
and $Q_{34}$. Then one has
\bea
&& \sqrt{|Q_{12} Q_{34}|} \ge  {1\over 2\pi}\int^{2\pi}_0 d\theta
\sqrt{|g_{55}|} \sqrt{|h_{12} h_{34}|} \ge
 {1\over 2\pi}\int^{2\pi}_0 d\theta
\sqrt{|g_{55}|} \sqrt{|\e^{ijkl} h_{ij}h_{kl}/8|} \nonumber\\
&& \ \ \ \ \  \ \ \ \ \ \ \ \ \ \ \  \ge (\sqrt{g_{(4)}})^{1\over 2}
{1\over 2\pi}\int^{2\pi}_0 d\theta
\sqrt{|g_{55}|}.
\ena
Therefore the arbitrary curve is constrained by the
conserved quantities as
\bea
|J'_{12}|+|J'_{34}| \le \sqrt{g_{(4)}} L^2 \le W(Q)\,.
\ena
This describes the moduli fluctuation of the bosonic degrees
with conserved charges and angular momentum.

For the case of $T^6$, one has 3 kind of simple holomorphic
four cycles defined by $Z^\alpha=C^\alpha$
($\alpha=1,2,3$)
and let us consider the case where m5
wraps each four cycle once. Then for each four cycle,
we get the above inequality. Hence we get
\bea
|J'_{12}|+|J'_{34}| \le \sqrt{g_{(4)}} L^2
\le {\rm Min}[W(Q_{(1)}), W(Q_{(2)}),
W(Q_{(3)})]\,,
\ena
where $I=1,2,3$ label the three holomorphic four cycles. We get the
further restriction because each cycle puts independent restrictions.

Finally the total energy of the system for the supertubes is given by
\bea
E= -\int^{2\pi}_0 d\theta \sqrt{|g_{55}|}\int_P h\wedge F,
\ena
where $P$ denotes the holomorphic four submanifold.

\section{M-theory on a CY space}

In this section, we would like to consider the embedding
of m5 and M2 system into a CY 3-fold
instead of $T^6$. As before, the m5-branes wrap four cycles
of the CY and the remaining one direction forms an arbitrary
curve in $789\natural$ directions which is again taken to be flat.

\subsection{Killing spinor}

Before describing the m5-brane problem, let us first describe the
Killing spinor of M-theory CY compactification.  In this
compactification, the remaining Killing spinor has eight real
components corresponding to $N=2$ supersymmetries. To describe
the nature of this spinor, we start from 32 component complex
spinor which has twice as many components of the maximal Killing
spinor in eleven dimensions. Further we introduce a vielbein
for the CY space whose only  purely holomorphic or antiholomorphic
components are  nonvanishing. This gauge choice of vielbein is
possible whenever the space is hermitian. The CY metric is then
given by $g_{\alpha\bar\beta}= E_\alpha^\gamma E^{\bar\gamma}_{\bar\beta}$
where the vielbein $E^\beta_\alpha$ satisfies
$E^{\bar\beta}_{\bar\alpha}=(E^\beta_\alpha)^*$ with $\alpha,\beta,\gamma
=1,2,3$. The eleven-dimensional gamma matrices are taken to be pure
imaginary.

The remaining component of the Killing spinor is constructed as follows.
Consider a spinor satisfying the projection
\bea
\tilde\Gamma_{\bar\alpha}\epsilon_+ =0, \ \ \ \ \  {\rm for}\  \alpha=1,2,3,
\ena
where we define $\epsilon_\pm$ by the projection
\bea
\Gamma_{(7)} \epsilon_\pm = \pm \epsilon_\pm\,,
\ena
with $\Gamma_{(7)}= i\Gamma_1 \Gamma_2 \cdots \Gamma_6$.

The gamma matrices  $\tilde{\Gamma}_{\bar{\alpha}}$ is defined by
${1\over 2}(\Gamma_{2\alpha-1}+i \Gamma_{2\alpha})$ whereas
$\tilde{\Gamma}_{\alpha}$ by
${1\over 2}(\Gamma_{2\alpha-1}-i \Gamma_{2\alpha})$.
At this stage, the total 32 complex space is projected by the factor of
$1/4$. We then impose the two conditions
\bea
&& \tilde{\Gamma}_\alpha \tilde{\Gamma}_\beta \tilde{\Gamma}_\gamma
\epsilon_+ =\epsilon_{\alpha\beta\gamma} \epsilon_- \,,\nonumber\\
&&\epsilon_-= \epsilon^*_+\,,
\ena
where the first choice relies on the existence of the nowhere vanishing
holomorphic $(3,0)$ form. By these two conditions, the space of spinor is
further projected down by the factor of $1/4$. The real eight-dimensional
spinor is constructed by
\bea
\epsilon_{CY}= c \epsilon_+ + c^* \epsilon_- ,
\ena
where $c$ is an arbitrary constant. This is the Killing spinor
we use for the study of BPS configuration of the CY compactification.

The induced vielbein on the worldvolume
\bea
e_i^a = \frac{\pa Y^b}{\pa \s^i} E^a_b,
\ena
relates the gamma matrices as
\bea
\c_i = e_i^a \Gamma_a = \pa_i Y^a E_a^b \Gamma_b, \quad
\c_0 = \Gamma_0, \quad
\c_5= X^m{}' \Gamma_m,
\ena
and the induced metric
\bea
g_{00} =1, \quad
g_{ij} = \pa_i Y^a \pa_j Y^b g_{ab}, \quad
g_{55}=X^m{}' X_m'.
\ena
Then the kappa symmetry condition is given by the same expressions
as those in  (\ref{kappa}) and (\ref{kappa1}).

The analysis of the BPS equations of section 2 goes through
if we choose a projection
\bea
\Gamma_0 \epsilon_+ = -\epsilon_+ ,
\ena
which leads to $P_1\epsilon_{CY}=
P_2\epsilon_{CY}=P_3 \epsilon_{CY}=\epsilon_{CY}$.
Further half of the supersymmetries are projected out
and only four real supersymmetries remain unbroken.
The BPS equations
then become those in (\ref{bps0}) -- (\ref{bps6}).

\subsection{Analysis of the BPS equations}

To solve our BPS equations, we  make again further gauge choice
\bea
(Y^1, Y^2, Y^3, Y^4) = ( \s^1, \s^2, \s^3, \s^4).
\ena
and introduce
the complex coordinate
\bea
Z=Y^5+iY^6= Z^3; \quad
w_1= \s^1 +i \s^2=Z^1, \quad
w_2=\s^3 +i \s^4=Z^2,
\ena
Then eq.~(\ref{bps2}) may be written as
\bea
\e_{ab} f^{ab\bar\alpha\bar\beta}=0\,.
\label{bps23}
\ena
This equation is solved by any holomorphic function $Z$
satisfying
\bea
\pa_{\bar w_1} Z = \pa_{\bar w_2} Z = 0\,.
\label{holocon}
\ena
To show  this, let us solve eq.~(\ref{bps23}) for $\bar\alpha=2$ and
$\bar{\beta}=3$. The equation becomes
\bea
\e_{ijkl}\pa_i Z^\alpha \pa_j \bar{Z}^{\bar{\beta}}
\pa_k \bar{Z}^{\bar{\gamma}} \pa_l \bar{Z}^{\bar{\delta}}
E^1_\alpha E^{\bar{1}}_{\bar{\beta}} E^{\bar{2}}_{\bar{\gamma}}
E^{\bar{3}}_{\bar{\delta}}
=
\e_{ijkl}\pa_i Z^\alpha \pa_j \bar{Z}^{\bar{1}}
\pa_k \bar{Z}^{\bar{2}} \pa_l \bar{Z}^{\bar{3}}
{\rm det}(E^{\bar{\gamma}}_{\bar{\delta}}) E^{1}_{\alpha}=0\,.
\ena
Considering also  $\bar\alpha,\, \bar{\beta}=1,\,2$ and
 $\bar\alpha,\, \bar{\beta}=1,\,3$ and  noting
 ${\rm det}(E^{\bar{\gamma}}_{\bar{\delta}})\neq0$,
one
gets
\bea
\e_{ijkl}\pa_i Z^\alpha \pa_j \bar{Z}^{\bar{1}}
\pa_k \bar{Z}^{\bar{2}} \pa_l \bar{Z}^{\bar{3}}=0\,,
\ena
which leads to the holomorphicity condition  (\ref{holocon}).
Eq.~(\ref{bps5}) is again solved for
\bea
H^{w_1 w_2} = h_{w_1 w_2}=0,
\ena
and eqs.~(\ref{bps3}) and (\ref{bps4}) imply
\bea
H^{i5} = H^{05}=0\,.
\ena
Eq.~(\ref{bps1}) can be written as
\bea
{1\over 8} \sqrt{g_{(4)}}\e_{ijkl} H^{ij} H^{kl}=-1,
\ena
and the last BPS equation (\ref{bps6}) follows
automatically as before.  The constraint  equation (\ref{gauss})
is again solved by (\ref{gausssol}). Thus we are left with
the final equation
\bea
{1\over 8 \sqrt{g_{(4)}} }\,\,\e^{ijkl}\, h_{ij} h_{kl}=1,
\label{ins}
\ena
with the Bianchi identity $dh=0$. One of the trivial solution of
the equation is given by
\bea
h = ik,
\ena
where $k$ is the K\"ahler two form.

The most general solutions are described as follows.
m5 wraps holomorphic four cycles, which is classified by the
the Hodge number $h^{1,1} (M)$ of the CY manifold $M$.
Or the four cycles are described by codimension 1
hypersurface defined as a zero locus of
holomorphic function on $M$. This precisely corresponds to
divisor of the CY manifold $M$. Each section of
line bundle ${\cal L}$ on $M$ is in one-to-one
correspondence with  divisor $P$. The deformation space of
the divisor is given
by $2\big(h^0(M, {\cal L}(P) )-1\big)=2 h^{2,0}(P)$~\cite{Maldacena}.

For any such four cycles $P$ in $M$, one has to solve the nonlinear
instanton equation (\ref{ins}). The full nonlinear analysis of
the equations seems very complicated. But one may find the
dimension of the solution space by linearizing the equation
around the trivial solution $h= ik + \delta h$. $\delta h$
satisfies the Hermitian Yang-Mills equation.  The solutions are
harmonic $(1, 1)$ form on $P$ with one extra constraint
that the harmonic $(1,1)$ form should be anti-selfdual. The only
selfdual harmonic $(1,1)$ form is the K\"ahler two form. The number of
dimensions of anti-self-dual harmonic $(1,1)$ form is thus
$h_-^{1,1}(P)=h^{1,1} (P)-1$.

Thus for each cohomology class of $H^{1,1}(M)$, most general solutions
involves $2h^{2,0}(P)+ h^{1,1}(P)-1= b_2(P)-1$ arbitrary functions
of $\theta$ in addition to the arbitrary curve in $789\natural$
directions, where $b_2$ is the Betti number. The curve involves only
3  physical degrees of the transverse fluctuation
 because of the reparametrization invariance
along $\t$ directions. Therefore the total number of degrees involved
with the arbitrary function of $\theta$ is given by $b_2(P)+2$, which
is related to
the Euler number by
\bea
b_2(P)+2= \chi (P)+ 4 b_1(P).
\ena
If the divisor is very ample, one has $b_1(P)= b_1(M)$ and then
$b_2(P)+2= \chi (P)$.
(Here we use the property  $b_1(M)=0$ for a CY manifold of exactly SU(3)
holonomy and not its subgroup.)
The Euler number is evaluated as
\bea
\chi(P) = \int_M (P^3 + P c_2 (M)),
\ena
where $P$ is the $(1,1)$ form of $M$ dual to the divisor and $c_2$
is the second Chern class.

For the divisor $P=T^4$ of $M= T^6$, the above $\chi(P)=0$ but
$b_1(P)=b_3 (P)=2$. Hence the number of moduli is 8 for the $T^4$;
three for the transverse fluctuation in the
$789\natural$ space, two for the $T^4$ moduli in $T^6$
and three for the flux moduli as we verified explicitly.

\section{Supertubes with 4 M2-Charges}

In this section, we consider the M-theory compactified on $T^8$
(or CY 4-fold) as a generalization of the 3 M2-charge case.
In this compactification,
we consider the configuration involving four M2-branes
extended along $(12)$, $(34)$, $(56)$, and  $(78)$ spatial directions
of the tangent space. This is what we call 4-charge supertubes in M-theory.

The setting to obtain the BPS equations from
(\ref{first}) to (\ref{f4}) are unchanged with a few modifications.
The indices $a,b,c$ run now from 1 to 8 and
$m,n$ stand for 9 and $\natural$.
The relevant projection operator becomes now
\begin{equation}
\epsilon = P_1 P_2 P_3 P_4  \epsilon_0,
\end{equation}
where $P_{1,2,3}$ are the same as before
and $P_4 = \frac{1 + i\Gamma_{078}}{2}$.

With the gauge choice $a
= t$, $H^{i0}=H^{50}=0$ automatically follow from the definition (\ref{hmn}).
The $\kappa$-symmetry condition gives
\bea
\label{bps3cahrge}
&& \Gamma \epsilon = \frac{\sqrt{-g}}{\sqrt{-(g+iH)}} \Big(
\frac{1}{\sqrt{-g}}\Gamma_0 \gamma_5 \Gamma_{abcd}f^{abcd}+
\frac{1}{2}H^{ij}\gamma_{ij}\Gamma_0 + H^{i5}\gamma_i \gamma_5
\Gamma_0 \nonumber\\
&&\ \ \ \  -\frac{1}{8}\sqrt{-g}\epsilon_{ijkl}
H^{ij}H^{kl}\gamma^5 \Gamma_0
-\frac{1}{2}\sqrt{-g}\epsilon_{ijkl}H^{5i}H^{jk}\gamma^l \Gamma_0
 \Big)\epsilon = - \epsilon \,.
\ena
The  3rd term in the big parenthesis  cannot be canceled
by others and, consequently, $H^{i5}=0$.

With the help of the projection operators,
the first term in the big parenthesis  may be arranged  as
\bea
&& \Gamma_{abcd}f^{abcd}\epsilon =
3 f^{ \bar{\alpha}\bar{\beta}\gamma \delta}
\delta_{\bar{\alpha}\delta} \delta_{\bar{\beta}\gamma} \epsilon
+
(f^{\alpha \beta \gamma
\delta }\tilde{\Gamma}_{\alpha}\tilde{\Gamma}_{\beta}\tilde{\Gamma}_{\gamma}
\tilde{\Gamma}_{\delta} -6f^{\bar{\alpha} \beta \gamma
\delta}\delta_{\bar{\alpha}\beta}\tilde{\Gamma}_\gamma
\tilde{\Gamma}_\delta )\eta_+ \nonumber\\
&&\ \ \ \ \ \ \ \ \ \ \ \ \ \ \ + ( f^{\bar{\alpha} \bar{\beta}
\bar{\gamma} \bar{\delta} } \tilde{\Gamma}_{\bar{\alpha}}
\tilde{\Gamma}_{\bar{\beta}} \tilde{\Gamma}_{\bar{\gamma}}
\tilde{\Gamma}_{\bar{\delta}} -6f^{ \alpha \bar{\beta} \bar{\gamma}
\bar{\delta}} \delta_{\bar{\beta}\alpha}
\tilde{\Gamma}_{\bar{\gamma}} \tilde{\Gamma}_{\bar{\delta}} )
\eta_- \,\,,
\ena
where $\eta_{\pm} = \frac{1\pm \Gamma_0}{2} \epsilon$, and $\alpha,
\beta,\gamma,\delta $ are holomorphic  indices running from 1 to 4.
Thus one is lead to the BPS equations,
\begin{equation}
f^{\alpha \beta \gamma \delta }\epsilon_{\alpha \beta \gamma \delta
}=0 ~~,~~f^{\bar{\alpha} \beta \gamma
\delta}\delta_{\bar{\alpha}\beta} = 0\,.
\end{equation}
The holomorphic target space coordinates are given by $Z^\alpha =
Y^{2\alpha -1}+iY^{2\alpha}$.  Again we choose the gauge $ w_1
=\sigma^1 + i \sigma^2=Z^1$ and $ w_2 =\sigma^3 + i \sigma^4=Z^2$.
Then the above equation implies simply that
$Z^3$ and $Z^4$ are holomorphic. Namely m5 is wrapping the holomorphic
four cycles.

The coefficient of $\gamma_5 \Gamma_0$ has to be zero. After some algebra,
this leads to the BPS equation
\bea
{1\over 8} \sqrt{g_{(4)}} \epsilon_{ijkl} H^{ij}H^{kl}=-1\,,
\label{nins1}
\ena
which agrees with the expression of the 3-charge case.

{}From the second term in the big parenthesis, one gets the
BPS equation
\bea
H^{w_2 w_2}=H^{\bar{w}_2 \bar{w}_2}
=0\,.
\ena

Then the final remaining equation requires
\begin{equation}
g_{p\bar{q}}H^{p\bar{q}}
=\frac{\sqrt{-(g+iH)}}{\sqrt{-g}}\,,
\end{equation}
where $p,q$ ($\bar{p},\bar{q}$) are the worldvolume
(anti) holomorphic indices running over 1,2. This last condition
again automatically follows from the above BPS equations.

Hence the m5 are again described by the holomorphic four
cycles of the eight manifold $M$. Then the worldvolume flux
of $(1,1)$ type satisfying (\ref{nins1}) may be turned on.

\section{Conclusion}

In this paper we have first discussed the 3-charge
supertubes in the M-theory compactified on M (that may be
 $T^6$ or CY 3-fold). The system under consideration consists
of M2 branes extended along the $(12)$, $(34)$ and $(56)$
tangent space directions of the internal manifold $M$.
The four spatial directions of m5-branes wraps four cycles of
$M$ and the remaining spatial direction forms an arbitrary curve
in the space transverse to the internal manifold.

By analyzing the m5-brane action in detail, we have obtained
the BPS equations governing the dynamics of the supertubes. The
nature of their solutions are fully understood. For some
particular cases,
we have worked out the solutions explicitly.

The most general supertube solutions involving 3 charges
may be described as follows. The four directions of m5
may wrap any holomorphic four cycle $P$, which is described by the zero
locus  of the holomorphic function over $M$. The holomorphic four
cycles are
dual to (1,1) type harmonic two forms restricted to the K\"ahler cone.
As said in the above, the remaining spatial direction forms an arbitrary
curve in the transverse space.
The worldvolume flux may be turned on, which has to be (1,1) type in $P$.
This two form flux has to satisfy the nonlinear instanton equation.

The $P$ may be deformed smoothly inside $M$ and the dimension of
this deformation space is given by $2 h^{2,0}(P)$. And also the
solution space dimensions of the instanton equation of the two form
are shown to be given by $h^{1,1}(P)-1$. Then the shape of $P$ and
the two form flux may vary over the moduli space as we move along
the curve directions as a function of $\theta$.

With fixed energy, angular momentum and other conserved
charges, the number of wraps of the cycles by m5 may be
arbitrary as the m5-brane charges are not conserved due to its
dipole nature. However, by the continuity of the deformation along the
curve direction, the number of wraps cannot be changed along the
curve directions because the number is topological. In this sense,
we expect that, in the moduli space of supertubes,
the number of wraps of $P$ labels superselection sectors of the
whole moduli space.

We have also included brief discussion of the 4-charge
generalization.

After analyzing the most general configuration, the natural question
is on the degeneracy of the configurations with fixed conserved
charges. This investigation will lead to the microscopic understanding
of the entropy of supertubes.
Since the system has an infinite number of degrees classically,
one has to quantize the moduli fluctuations properly. The
contribution of the fermions should also be included if there are any.
In fact this problem has been considered in Ref.~\cite{stro} but we would
like to be more explicit and fill some possible gaps.
This will be our main subject of the further investigations.

\vskip 1cm

\section*{Acknowledgment}
We would like to thank A. Tsuchiya for valuable discussions
at the early stage of this work.
D.B. is supported in part
by KOSEF ABRL R14-2003-012-01002-0, KOSEF R01-2003-000-10319-0
and KOSEF SRC CQUeST R11-2005-021.
The work of NO was supported in part by the Grant-in-Aid for
Scientific Research Fund of the JSPS No. 16540250.

\vskip 1.5cm

\newcommand{\NP}[1]{Nucl.\ Phys.\ B\ {\bf #1}}
\newcommand{\PL}[1]{Phys.\ Lett.\ B\ {\bf #1}}
\newcommand{\CQG}[1]{Class.\ Quant.\ Grav.\ {\bf #1}}
\newcommand{\CMP}[1]{Comm.\ Math.\ Phys.\ {\bf #1}}
\newcommand{\IJMP}[1]{Int.\ Jour.\ Mod.\ Phys.\ {\bf #1}}
\newcommand{\JHEP}[1]{JHEP\ {\bf #1}}
\newcommand{\PR}[1]{Phys.\ Rev.\ D\ {\bf #1}}
\newcommand{\PRL}[1]{Phys.\ Rev.\ Lett.\ {\bf #1}}
\newcommand{\PRE}[1]{Phys.\ Rep.\ {\bf #1}}
\newcommand{\PTP}[1]{Prog.\ Theor.\ Phys.\ {\bf #1}}
\newcommand{\PTPS}[1]{Prog.\ Theor.\ Phys.\ Suppl.\ {\bf #1}}
\newcommand{\MPL}[1]{Mod.\ Phys.\ Lett.\ {\bf #1}}
\newcommand{\JP}[1]{Jour.\ Phys.\ {\bf #1}}

\end{document}